\documentclass[preprint,12pt]{elsarticle}
 \usepackage{graphics}
\usepackage{amssymb}
\journal{Journal of Crystal Growth}
\begin{document}
\begin{frontmatter}
\title{Long time evolution of meandered steps during the crystal growth processes}
\author{Magdalena A. Za{\l}uska--Kotur }	
\ead{zalum@ifpan.edu.pl}
 \address{Institute of Physics, Polish Academy of Sciences,
Al. Lotnik{\'o}w 32/46, 02-668 Warsaw, Poland and Faculty of Mathematics and Natural Sciences,
Card. Stefan Wyszynski University, ul Dewajtis 5, 01-815 Warsaw, Poland}
 \author{Filip Krzy{\.z}ewski} 
\ead{fkrzy@ifpan.edu.pl} 
 \address{Institute of Physics, Polish Academy of Sciences,
Al. Lotnik{\'o}w 32/46, 02-668 Warsaw, Poland} 
\author{Stanis{\l}aw Krukowski} 
\ead{stach@unipress.com.pl}
\address{ High Pressure Research Center, Polish Academy of Sciences, ul. Soko\l owska 29/37,01-142 Warsaw, Poland and
Interdisciplinary Centre for Materials Modeling, Warsaw University, Pawi\'nskiego 5a, 02-106 Warsaw, Poland }
\begin{abstract}
Step meandering during growth of gallium nitride (0001) surface  is studied using kinetic Monte Carlo method. Simulated growth process, conducted in N-rich conditions are therefore controlled by Ga atoms surface diffusion. The model employs dominating four-body interactions of Ga atoms that cause step flow anisotropy during growth. Overall kinetics and shape selection features of step meandering are analyzed assuming their dependence on the external particle flux and on the temperature. It appears that at relatively high temperatures and low  fluxes steps move regularly preserving their initial shapes of straight, parallel lines. For higher fluxes and at wide range of temperatures step meandering happens. It is shown that, depending on the initial surface parameters, two different scenarios of step meandering are realized. In both these regimes meandering has different character as a function of time. For relatively high fluxes meanders have wavelengths shorter than the terrace width and they grow independently. Eventually surface ends up as a rough structure. When flux is lower regular pattern of meanders emerges. Step meander even if Schwoebel barrier is absent. Too high barrier destroys step stability. The amplitude of wavelike step meanders increases in time up to a saturation value. The mechanism of such meander development  is elucidated. 
\end{abstract}

\begin{keyword}
A2 growth from vapor \sep A1 computer simulation  \sep A1 surface processes\sep A1 surface structure 
\end{keyword}
\end{frontmatter}
\section{Introduction}
\label{sec:A}
       Crystal growth dynamics and its relation to the  formation of various geometric patterns remains a subject of continuous interests of many researchers. At its inception as scientific notion, crystal growth geometry was supposed to follow simple scheme devised by Burton, Cabrera and Frank (BCF) \cite{[1]}. Flow of straight parallel steps, alimented by the diffusion from the terraces, being at the core of terrace-ledge-kink (TLK) model, was supposed to sustain the growth uniformly as long as the steps were at abundance \cite{[1]}. This scenario was confirmed by early Monte Carlo simulations in the 70ies. For small systems, of the size of 20 lattice constants, it was confirmed that the overall picture of the step motion and their dynamics follows BCF step flow model \cite{[2],[3],[4],[5]}.  

       This simple picture was enriched by additional phenomena either emerging during Monte Carlo simulations or observed in the real growth experiments. It was noticed that parallel steps have tendency to create step trains, in accordance to the prediction of kinematic step train theory, proposed first by Frank and later discuessed by Vekilov et al. \cite{[8]}. This tendency could lead to a creation of train of steps , double steps, macrosteps of even formation of new crystallographic face during growth \cite{[9],[10],[11],[12],[13]}. The existence of such features were explained either by invoking the impurities or inclusions of foreign phase \cite{[14],[15]} or by coupling between the transport and step motion, leading to the spatial nonuniformity and step accumulations at some regions \cite{[16]}.

       In parallel to the step motion instabilities, triggered by external factors, as above mentioned supersaturation or impurities, the inherent factors were also identified. The structure of the steps, first assumed to be uniformly microscopically rough, could be affected by long range fluctuations. The analogy between the discrete Gaussian model and two-dimensional Coulomb gas was used by Chui and Weeks \cite{[21]} and H. Muller-Krumbhaar to demonstrate that interface widths diverge with the system size as $L^{-1/2}$ in 2D and as $\ln L$ in 3D \cite{[22]}. Thus the step width, i.e. the edge of two dimensional system, diverges with the size as $L^{-1/2}$   \cite{[23]}. Qualitatively, such increase of the long range fluctuations with the size of 2D system was observed first  in 2D Monte Carlo simulations by Krukowski and Rosenberger \cite{[24]}. 
      
       Similar instability phenomena, denoted as step meandering, were observed during growth of various type crystals \cite{[25],[26], maroutian, hibino}. The experimental data include step meandering in various systems, both metallic and semiconductor. Step meandering was observed during growth on  vicinal surfaces, such as Cu(1 1 17) or Cu(0 2 24) \cite{maroutian}. Similarly, step meandering was identified during growth of silicon layers. Noteworthy, step meandering was observed during growth on densely packed Si(111) surface \cite{hibino,omi}. Such phenomenon was descerned on vicinal Si(100) surface \cite{pascale}. It was also shown that electromigration of adsorbed atoms considerably enhances strength of this phenomenon \cite{leroy}. The step meandering, and other morphological instabilities on silicon were exhaustively discussed by Yagi et al. \cite{[26]}
       
       In addition, in Monte Carlo simulations such behavior was also observed \cite{[27]}. Therefore, step meandering was recognized as important phenomenon which triggered activity in this domain. Most notably, several analytical approximate approaches were proposed, including Kardar, Parisi, Zhang (KPZ) equation, describing kinetic roughening of the step during growth of epitaxial layers\cite{[28]}. The KPZ equation, derived using symmetry arguments, is frequently used to analyze scaling properties of the Eden model describing growth of epitaxial layers in low temperature Molecular Beam Epitaxy (MBE) processes. Another approach, based on more physical arguments was undertaken by Bales and Zangwill who considered linear stability analysis of single step with respect to meandering \cite{[29]}. They showed that the instability is related to different incorporation rates from the lower and upper terraces and is diffusional in nature. Their analysis was extended to the nonlinear case by Bena et al. who derived equation governing the temporal evolution of the system \cite{[30]}. The equation was in fact published earlier by several authors \cite{[31],[32],[33]}, and is therefore somewhat mistakenly referred as Kuramoto-Shivashinsky equation \cite{[23]}. Bena et al showed that the instability is followed by chaotic behavior of the system \cite{[30]}. 
 
       These analytical approaches are able to provide trends and scaling behavior of the meandering step system. Yet the critical assessment of these theories requires careful analysis of the growth experiments, and parallel Monte Carlo simulations. Therefore both approaches are needed. Recently we have developed the model which was able to recover basic features of the growth of GaN layers by metaloorganic vapor phase epitaxy (MOVPE). It was shown that the step anisotropy could be obtained within model based on four-body interaction\cite{[34a],[34]}. It was also shown that in addition to step pairing, the step meandering could be obtained for large system sizes, which is also observed in real crystal growth experiments \cite{[35]}. The model is used below in large scale extensive kinetic Monte Carlo simulation allowing to verify basic features of the step meandering during growth of gallium nitride layers. 

In this work we show that meandering happens also in systems of symmetric steps without any Schwoebel barrier. Linear stability analysis approach does not predict such behavior. Such analysis usually does not takes into account step flow phenomenon. Our results show that step flow can be the only source of step instability. We study late stages of the meandering process, and observe that for low fluxes  meander amplitudes eventually saturate and regular step structures build up at the crystal surface. 

\section{The model}
\label{sec:B}
       \begin{figure}
\includegraphics[width=6cm,angle=0]{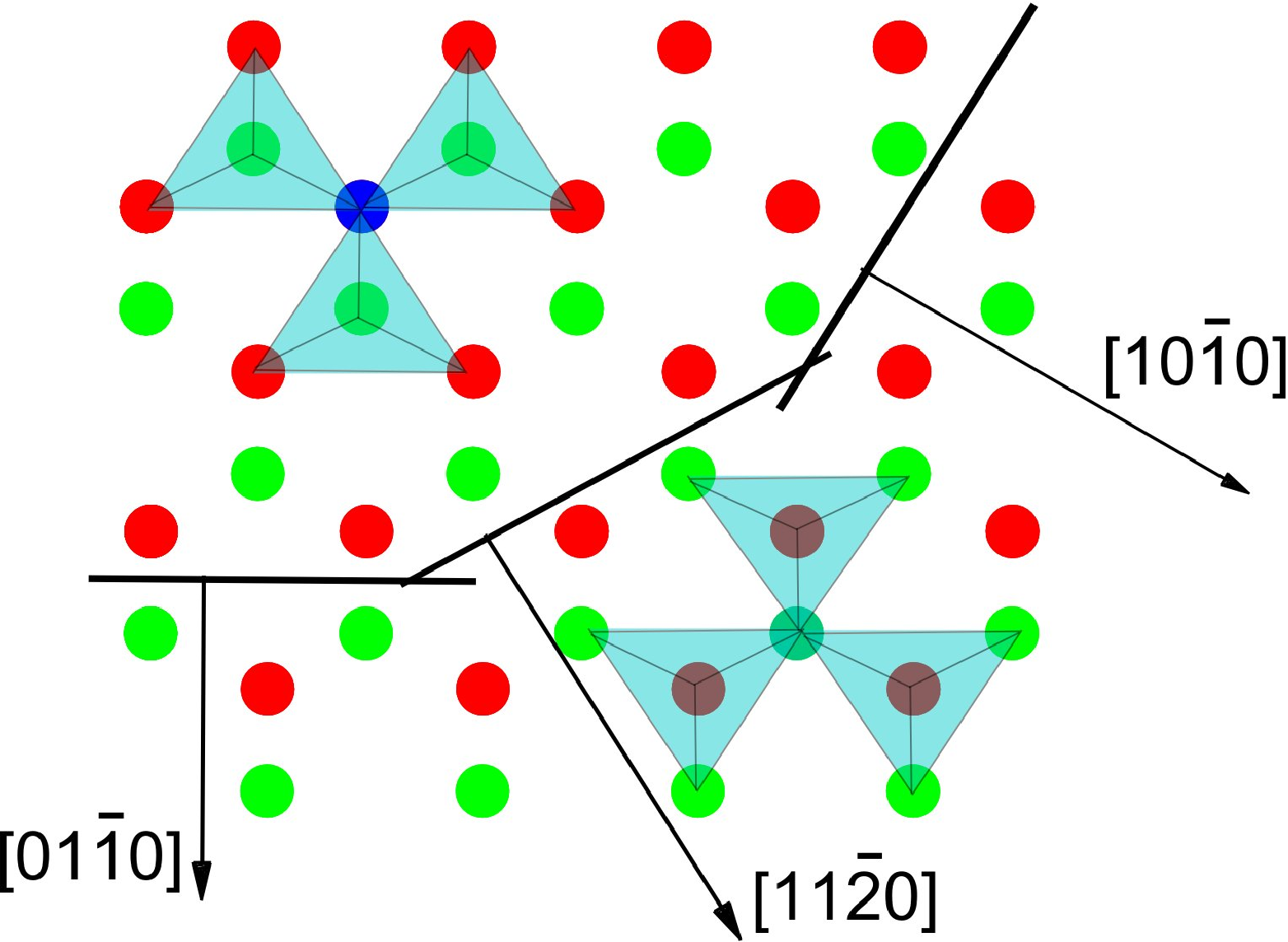}
\includegraphics[width=6cm,angle=0]{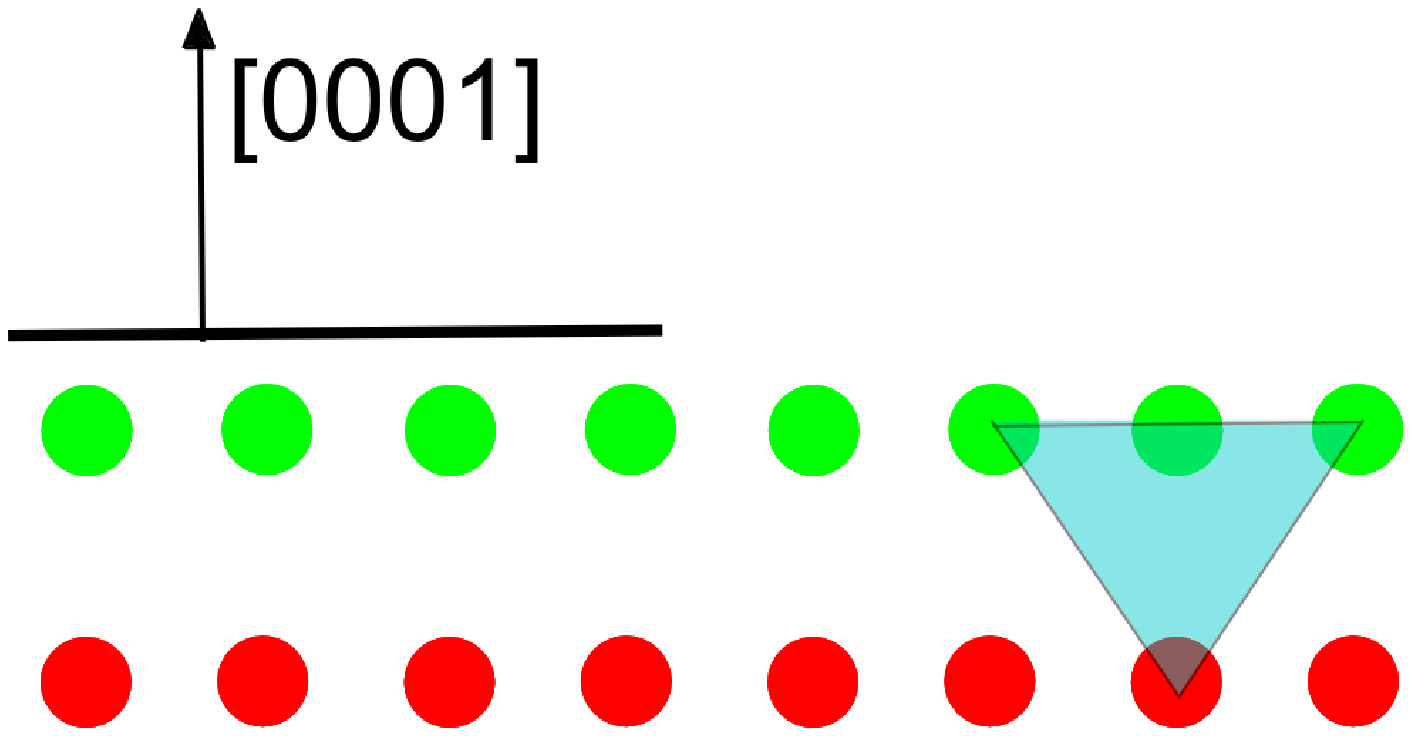}
\caption{\label{model} (color online) Model of GaN crystal. Ga atoms at two  layers are plotted by red and green  circles. Terrace height decreases by one layer at step marked by black line. At upper terrace red particles build  top layer and at lower terrace top layer consists of green particles.  Each tetrahedron contains one N atom inside. Two different bond orientations at subsequent terraces are marked by blue triangles.  }
\end{figure}

       We simulate growth of GaN(0001) layer in the N rich conditions. It is controlled by Ga atoms accumulated and diffused on top of growing crystal. We model the wurtzite lattice of the GaN crystal out of tetrahedrons of Ga atoms centered on N atom as illustrated in Fig 1. Three bonds connecting Ga atom with the nearest N atoms within one layer are rotated by 60$^o$ with respect to the bonds in the consecutive Ga layer. As a consequence of this crystallographic change step changes its character on coming one terrace down or on changing its orientation by 60$^o$. 
We have shown in the  previous work  \cite{[34]} that  the simplest way to account for the difference of two consecutive layers is to introduce four-body interaction between Ga atoms. 
Accordingly  our system is described by modeling the energy and dynamics of Ga atoms only. We assume that the energy  affecting the jump probability of each Ga from the lattice site depends on the number of Ga neighbors and on the N position as a tetrahedron center. It is determined in the following way:
\begin{equation}
\label{en_od_trojki}
n_i=\left\{\begin{array}{ll}
1,\quad \textrm{when tetrahedron has all atoms;} \\
\frac{1}{3} r\eta,\quad \textrm{when tetrahedron has empty sites,} \\
\end{array} \right.
\end{equation}
where $\eta$ is a number of occupied neighboring sites, belonging to  a selected tetrahedron and $r$ describes the relative  strength of the four-body and the two-body interactions in the system. When $r=1$ two body Ga-Ga interactions sum up to the value characteristic for fully occupied tetrahedron i.e. no additional four-body Ga interactions are present in the system. When $r<1$ three pair bonds to the nearest neighbors of a given particle in tetrahedron do not sum up to a value of one multiparticle bond. In such a case tetrahedron energy is not a simple sum of two-body interactions. This is the case we study below, the value $r=0.4$ is used throughout remainder of this work. It should be noted that the main results recounting step meandering process do not depend on the parameter $r$ significantly.

At GaN(0001) surface, each Ga surface atom  belongs potentially to four tetrahedrons, three in the  present layer and the one above. Its total energy could be expressed as
\begin{equation}
\label{en_czastki}
\alpha(J)=J\sum_{i=1}^{4}n_i.
\end{equation}
where parameter $J$ scales bonding energy, and the sum runs over four tetrahedrons, that surround every atom. 
 
In the study it is assumed that GaN crystal growth is controlled by kinetics of Ga atoms. It is typical, that  the differences of the concentration of Ga transporting agent in the neighboring vapor are negligible over the distances comparable to the typical terrace widths, therefore Ga atoms are adsorbed at the surface uniformly. Thus the Ga adsorption is accounted for by creation of an adatom at any empty adsorption site, at each MC step, with a rate used for modeling external particle flux
\begin{equation}
\label{F}
F=\nu_a e^{-\beta\mu},
\end{equation}
 where $\mu$ is a chemical potential, $\nu_a=2$ sets the timescale of simulation and $\beta=1/k_BT$ with $T$ as the temperature and $k_B$  Boltzmann constant. Each adsorbed particle diffuses  over the terrace  until it is attached at the steps. Thus, the possibility of reevaporation is neglected. Probability of a jump from the initial to the final site, in the diffusional movement, is given by diffusion parameter D, expressed as
\begin{equation}
\label{D}
D=D_0 e^{-\beta E},
\end{equation}
where $D_0=1$ is diffusion timescale, and 
\begin{equation}
E=E_B-\alpha_i(J)
\end{equation}
is the difference between the transition state energy $E_B$ and initial state bonding  energy $\alpha_i(J)$. Note that the uniform increase of the energy barrier for all jumps by the same value amounts to mere rescaling  of the timescale. We checked also system dynamics with additional barrier, which increased the difference between jumps along step and over terrace, but final structure of meanders did not change much except for relatively slower terrace diffusion.  We construct barrier for diffusion in such a way that it is lower for jumps along step than for jumps over terrace. The simplest way for construction of such a barrier is choosing  $E_B=\min[\alpha_f(J),\alpha_i(J)]$  where $\alpha_f$ is bonding energy of the final state of diffusing atom. As a result we get the following relation:
\begin{equation}
\label{eq:E}
E=\left\{\begin{array}{ll}
0,\quad \textrm{if $\alpha_i(J)<\alpha_f(J)$;} \\
\alpha_i(J)-\alpha_f(J),\quad \textrm{otherwise.} \\
\end{array} \right.
\end{equation}
which ensures lower energy barrier for jumps to the step and along step.
Adsorption rate at the step is additionally modified by Schwoebel barrier \cite{Schwoebel} that sets up different probability for atoms jumping to the step from upper and lower terraces. The height of the barrier is described by the parameter $B$. Probability of the jump over the barrier is as follows:
\begin{equation}
\label{DB}
D_B=e^{-\beta B} D.
\end{equation}
thus, the height of the barrier $B$ modifies jump rate given by equation (\ref{D}). As typically assumed, we impose Schwoebel barrier for the jumps from upper terrace. One barrier height  $B$ is used for each jump which crosses step. The difference in dynamics of both types of steps    is described by the step adsorption  rate which  varies with the step  type.

Crystal surface microstate is modeled by setting two uppermost layers of atoms. Every second layer of Ga atoms has different bond orientation. In all simulations the surface  is misoriented along one direction, coming up, i.e. when new step appears, the upper layer is converted into the lower one, and a new layer is built on top of the terrace. In such a way a continuity of particle-particle interaction at the step is guaranteed.   For the same density of adatoms, the step velocity depends on the particle interaction and therefore it changes with its location and orientation. 

Our simulations start with an even  number $n$ of equally spaced by $d$ lattice constants, straight steps. Heights of the neighboring steps differ by one Ga atomic layer. Periodic boundary conditions are applied in the lateral direction and in the direction in which the crystal grows they are helical, i.e. they are corrected by constant height difference between both ends of the system.

The simulation schema outline is as follows: new Ga atoms materialize at any of lattice site with probability (\ref{F}). If new adatom appears it  diffuses by jumps to the  nearest neighboring sites with  interaction dependent speed in accordance to (\ref{D}) and then the diffusive step attachment jump is modified by Schwoebel barrier (\ref{DB}). In such a way crystal growth is realized. Apart from the single particle events described above no other actions are realized during the simulation. 

\begin{figure}
\includegraphics[width=8cm,angle=0]{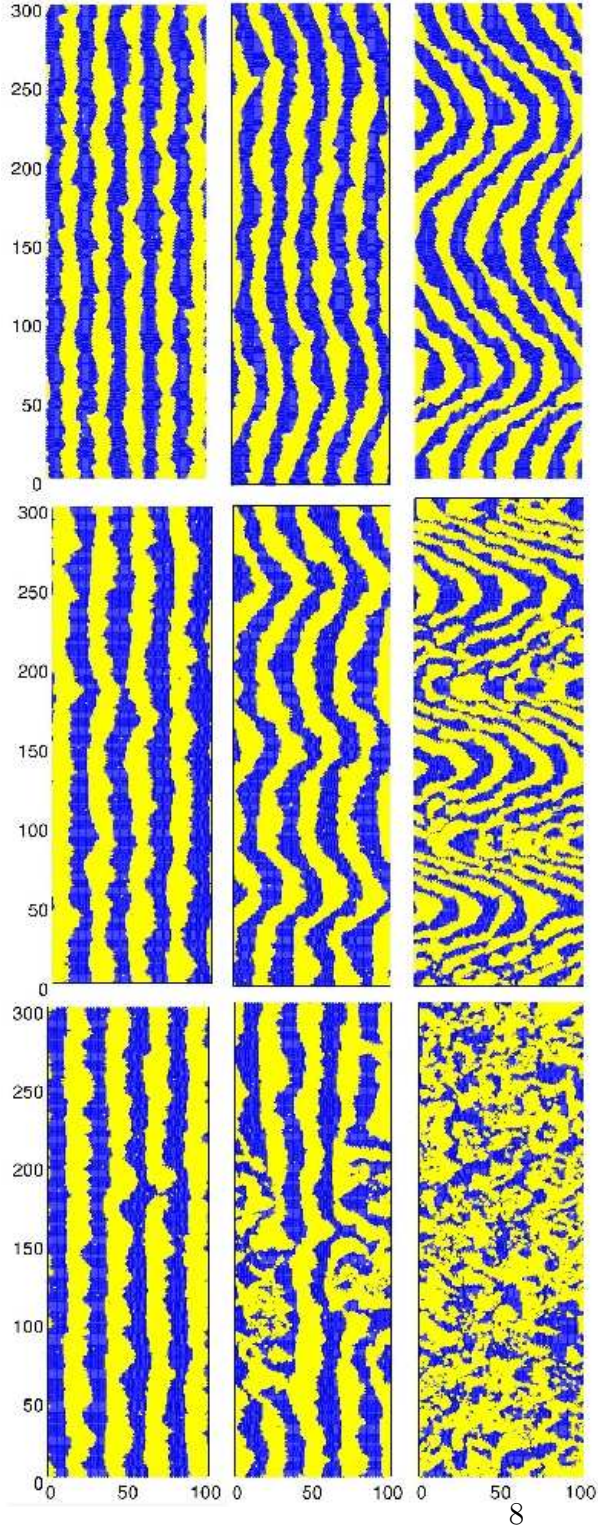}
\caption{(color online) Comparison of meandering pattern with zero, low and high Schwoebel barrier. Steps are   perpendicular to $[10\bar{1}0]$  and separated by $10 a$.    Simulation was carried out for system of size $300 \times 100$ lattice constants, $r=0.4$ , $\beta J=5.5$ and $\beta \mu=14.5$, at top row $\beta B=0$, at middle $\beta B = 0.5$ and bottom $\beta B= 1 $, respectively. For $B=0$ case plots present the pattern with 2 layers grown for $10^6$ MC steps, 45 layers for $4.5^. 10^7$ MC steps and $90 $ layers for $9^.10^7$ MC steps from left to right respectively. For both system below the patterns are shown of 2 layers ($10^6$ MC steps),  10 layers ($ 1^.10^7 $ MC steps) and the right with  50 layers ($ 5^.10^7$ MC steps) grown.  
}
 \end{figure}

\section{Creation of meandered structure.}
\label{sec:B2}
On studying surface evolution for various temperatures, fluxes and crystal miscuts we found set of parameters for which system builds up regular meander structure. This structure results in characteristic, stable during growth  pattern  of lines, oriented vertically to steps. We obtain such regular structures from the simulations of microscopic model. The interesting aspect of these results is that the asymmetry due to the step movement  is enough for the step to meander. To obtain this result we  do not need any additional income from the Schwoebel barrier asymmetry. It is illustrated in Fig 2. Top and middle rows of this figure compare step meandering with Schwoebel barrier absent in the model with the situation of relatively low barrier equal to half of the interaction strength. In both cases the same external flux was applied. It can be seen, that in the case of zero Schwoebel barrier we have longer meanders and it takes twice as time to build them. The step movement is neglected in most of analytical approaches \cite{[23],[29],[30]}, whereas as was shown in Ref. \cite{Dufay} step advection in fact cannot be ignored. This agrees with our observations. Up to some point step-flow advection during crystal growth affects the system in similar way as usually Schwoebel barrier does \cite{[23]}. Such correspondence, however, stops working when Schwoebel barrier is too high. As we can see in Fig 2, in the bottom row of pictures when Schwoebel barrier is too high, particles stick together breaking step continuity and disordered surface builds up. We  conclude that the regular step structures can be build at the surface only when steps have possibility of exchanging  particles. Regular, stable in time  meandered step structures  build up only for low enough fluxes. As we see below for higher  fluxes step meandering process has completely different character. Described above main features of meandering stay the same for $r=0.4$ and $r=1$. The only difference is in the relative width of the even and odd terraces at the left and right sides of meanders.
\begin{figure}
\includegraphics[width=10cm,angle=0]{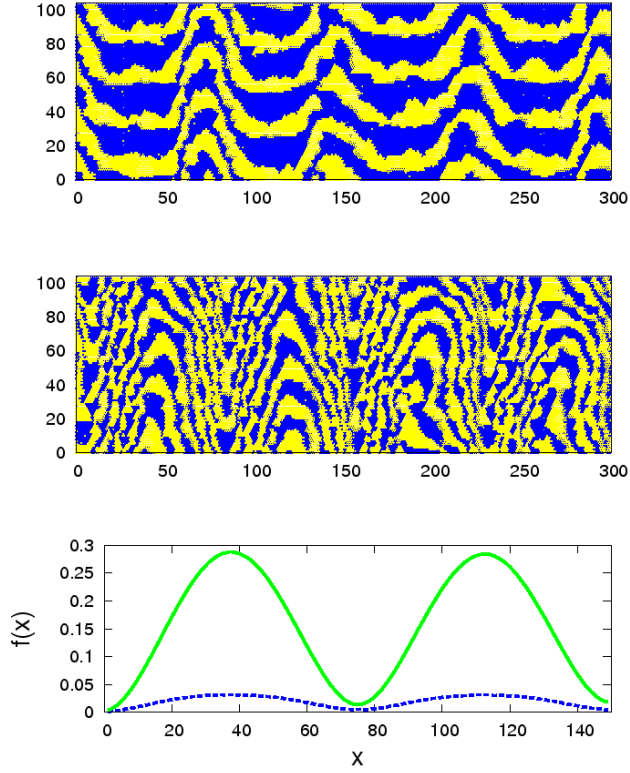}
\caption{\label{regular} (color online) Meandering pattern which evolves towards regular wavy picture is shown in the system growing slowly  with steps  initially oriented perpendicularly to $[10\bar{1}0]$  and separated by $10 a$.  At the bottom height correlation function versus distance along steps is plotted for both patterns shown above.  Simulation was carried out for system of size $300 \times 100$ lattice constants, $r=0.4$ , $\beta B=0.5$, $\beta J=5.5$ and $\beta  \mu=13.75$. Topmost plot presents situation with  20 layers ($10^7$ MC steps)  and middle one with 100 layers ($ 5^.10^7 $ MC steps),  grown. 
}
 \end{figure}
 	
\begin{figure}
\includegraphics[width=12cm,angle=0]{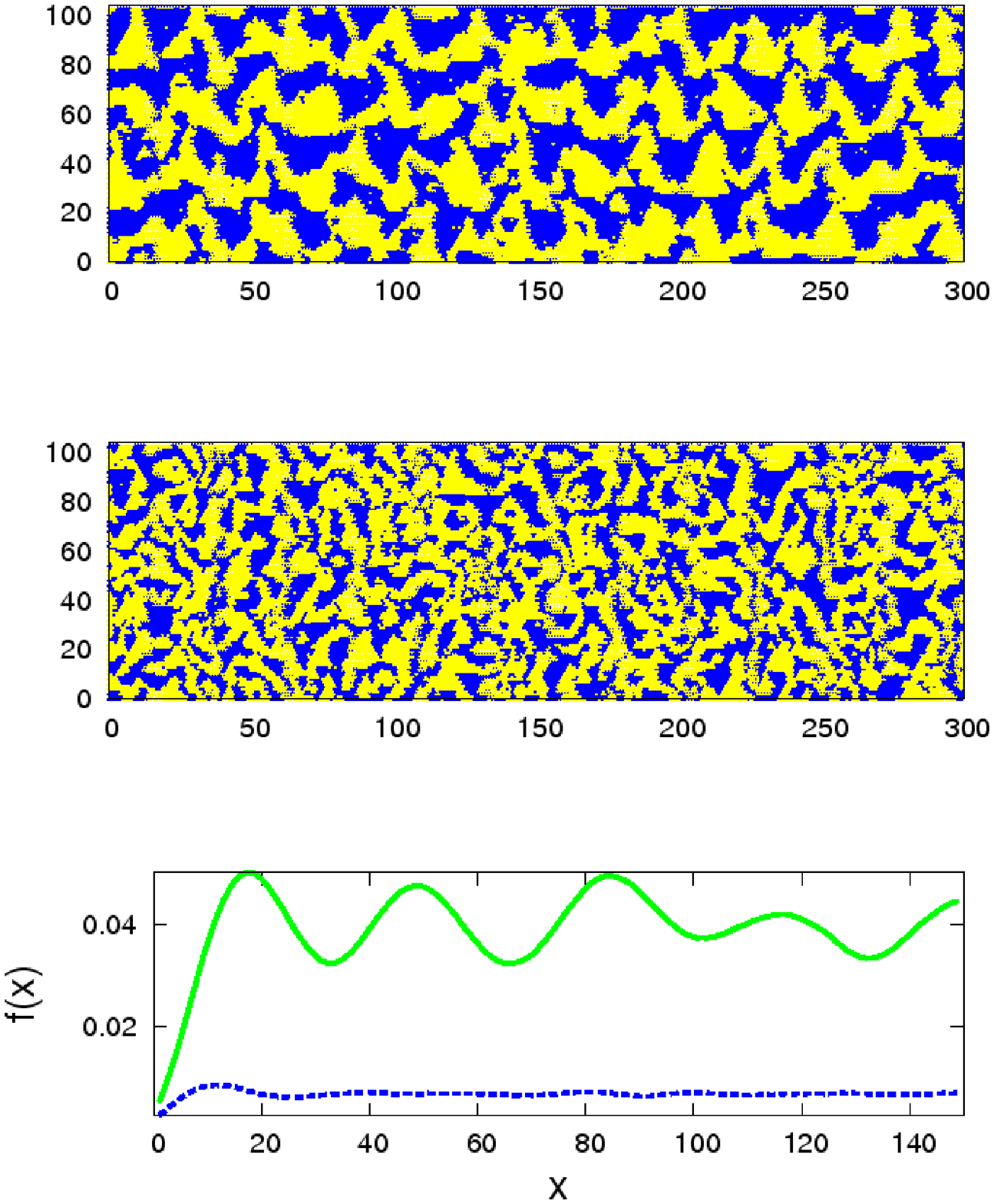}
\caption{\label{3d} (color online) Short wavelength pattern which evolves towards rough surface is shown in the system growing rapidly with steps  initially oriented perpendicularly to $[10\bar{1}0]$  and separated by $10 a$.  At the bottom height correlation function versus distance along steps is plotted for both patterns shown above.  Simulation was carried out for system of the size of $300 \times 100$ lattice constants, $r=0.4$ , $\beta  B=0.5$, $\beta  J=5.5$ and $\beta \mu=12.1$. Topmost plot presents situation with  3 layers ($2.5^.10^5$ MC steps) and middle one with 11 layers ($10^6$ MC steps)  grown. }
\end{figure}

During  the process of step meandering steps bend into the wavelike pattern, with the wavelength shorter for lower surface temperature and for smaller terrace width. Meanders which are long in comparison with the mean distance between steps create regular pattern of wavy like structure. Amplitude of these waves increases slowly during growth. Meanders of smaller wavelengths evolve in different way, becoming more and more irregular and finally form rough surface structure. We can compare  an example of regular pattern obtained for lower external flux, plotted in Fig. 3 and the rough surface, obtained for higher flux, shown in Fig. 4. The step height correlation function 
\begin{equation}
\label{f}
 f(x_1-x_2)=<[h(x_1) - h(x_2)]^2 >, 
\end{equation}    
calculated for distances measured along vertical, parallel to step direction, where $h(x)$ is  step height  at point $x$. Average $<>$ is taken over all system. It can be seen that the correlation function looks differently for these  two surface patterns.  When structure is well ordered, height correlation (\ref{f}) oscillates from zero to zero value. Such structure is shown in Fig. 3 for surface patterns after two different numbers of simulation steps. In both cases the height difference falls down to zero, repeating such behavior regularly at the distance equal to the meander wavelength. This does not occur for the case shown in  Fig. 4. The correlation function calculated for the system, not so well ordered, is small, with minute amplitude of its oscillations. Such structure of $f$ shows that some characteristic length exists in the system only locally, as can be seen from the top panel of Fig 4. , but it vanishes at larger distances.  For longer times such structure is concealed by the global step disorientation  as shown in the middle panel of Figure 4. 
\begin{figure}
\includegraphics[width=8cm,angle=-90]{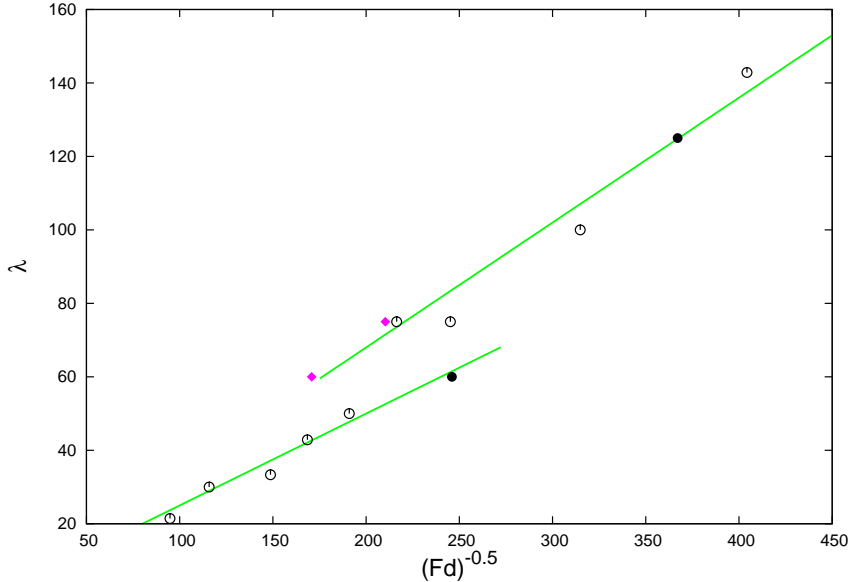}
\caption{\label{ph_di} (color online)
 Wavelength of the step meanders as a function of average particle flux arriving at single site of the step $Fd$ i.e. particle flux impinging single lattice site at the terrace from the vapor, multiplied by the terrace width. Data are shown for $\beta J=5.5$ and the following terrace widths: for d=10a are plotted using open circles,  for  d=20 - full black circles and for d=5 - purple diamonds. All data are located along the lines, fulfilling two different $\lambda\sim (Fd)^{-0.5}$ relations. The lower line describes  rough systems and the upper one - the systems which form regular patterns.
}
\end{figure}

	The characteristic wavelength of the step meandering can be measured as the  position of the first minimum of the function (\ref{f}). The length defined in such a way can be found in the regular, ordered system,  and also in the not so well ordered one, as these in Figs 3 and 4. Starting from the initial configuration of straight steps, the characteristic length of the step meanders grows rapidly up to its final value, for all studied cases.  The wavelength increase follows an universal function of the grown layer number $n$ and it can be written as $ \lambda \sim n^{0.2}$.  Such exponent is characteristic for early stages of pattern formation of the driven system \cite{levine}. It describes time growth of the characteristic length of domains in the direction perpendicular to drive. In our situation the system is driven by step movement and $\lambda$ is measured along direction perpendicular to the velocity.

 $\lambda$ for each system reaches  its plateau at  different values of $n$. The final wavelength is proportional to characteristic power $0.5$ of the flux. The lower flux and the shorter inter—step distances, the meander wavelength is longer, thus we get $\lambda \sim (Fd)^{-0.5} $. This dependence is illustrated in Fig. 5. This, quite general  relation, comes out from various types of calculations \cite{[23]} and appears to change when the diffusion along  the steps is seriously distorted \cite{nita}. In our case jumps are thermally activated with barrier given by site bonding energy calculated as inter--particle interaction. As long as the particle is attached to the step it wanders along it quickly and unhindered. It is stopped for some time at kinks, where its energy is lower. Time needed for particle to jump out of a kink changes with the temperature and that is why meanders have shorter wavelength at lower temperatures. 

Generally,  the wavelength of meanders is of the order of a distance passed by a particle along the step during the time separating two subsequent particles attaching to the same site at the step.  Assuming that particle realizes free random walk along the step and account that mean time between  particles striking given point of step can be obtained from the particle flux and inter—step distance, we obtain the following dependence: $\lambda ^2= \bar{D}a/Fd$. Note that the proposed relation, based on molecular processes, is similar to the criterion proposed by Krukowski and Tedenac in their analysis of step roughening phenomenon, leading to transition to fractal shape \cite {tedenac}. It is also worth noting that the relation proposed here is different from Bales Zangwill prediction, based on Mullins-like line diffusion, where the wavelength is inversely proportional to the terrace width \cite{maroutian,[29]}. If the diffusion $\bar D$ does not depend on $F$ or on $d$, the relation is straightforward. However, if $\bar D$ changes when $F$ is modified, the time dependence of $\lambda$ is not so simple. In Fig. 5 we present a single change of the type of the time dependence. This change can be related to the fact, that for relatively high flux, located at the left side of the plot, the particles attach to the step quite often, hence many kinks are present, which causes decrease of the mean diffusion constant. The lower line has smaller slope. On the other hand, the presence of many kinks leads to the surface roughening, hence all of the points along the lower line represent system, which eventually ended up as rough surface, like the one shown in Fig. 4. Additional experiments are needed to verify prediction based on different types of arguments. 

The upper line is plotted for the smooth meandering process, obtained for lower fluxes. These points also follow the $\lambda \sim (Fd)^{-0.5}$ dependence. For these points however, there is much more dispersion. They all are obtained for meanders of the length comparable or longer than the system size, hence boundary conditions become important. For wavelengths close to the system size one cannot be sure if this wave should not be slightly shorter or longer. Therefore it is very difficult to determine whether the increase of the meander length  breaks down rapidly or changes smoothly with the flux. In Fig. 5 we show data for one temperature and most of them were obtained for single terrace width $d=10$. We show however, that the points for terraces twice wider or narrower follow the same line. Transition from one line to another depends on the ratio $\lambda/d  $ and occurs for this value close to $2 \pi$, hence for each surface miscut it occurs for different $F d$ value.  The same wavelength that results in smooth meandering for small inter—step distance becomes too short for longer inter—step distance which leads to the rough surface pattern.  Hence  both points plotted for $d=5$ lay  at upper curve and two different points for $d=20$ are located at two different lines. 

Most examples we analyze here are for much narrower terraces, than those are observed experimentally however, as shown above, proper rescaling of the flux values is sufficient to  obtain the same behavior  for wider terraces.   
\begin{figure} 
\includegraphics[width=8cm,angle=-90]{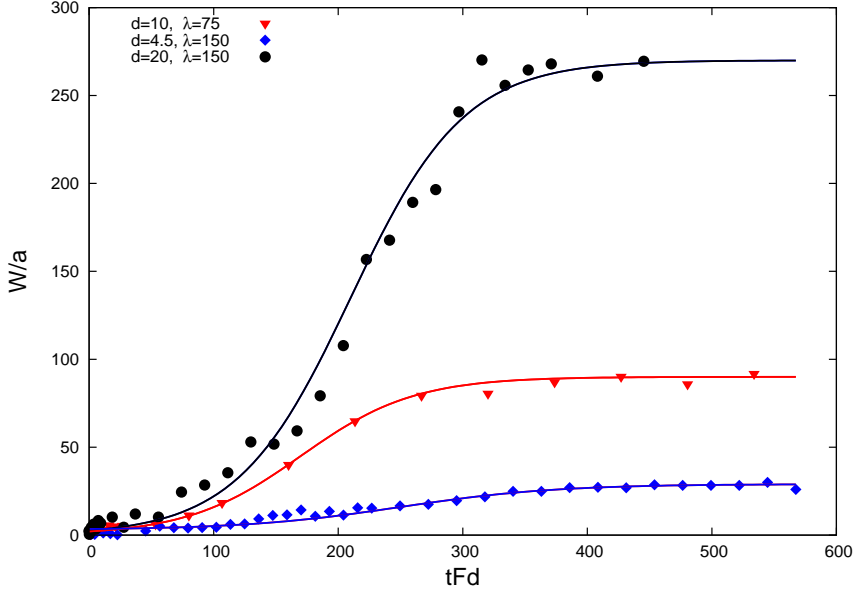}
\caption{(color online)
Time evolution of the mean width of step meanders for three different surface miscuts. All data are plotted for systems at temperature given by $\beta J=5.5$. }
\end{figure} 

\section{Time dependence of meander amplitude}
	
We measure the development of the meandering by determining the amplitude of step meander waves. The amplitude is calculated as a difference between the most advanced and the most retarded points of given step, measured along direction perpendicular to the step initial orientation. We average the amplitude value over all steps in the system and plot it as a function of the simulation time. For all studied cases the time evolution of step meander amplitude is very similar.  
In Figure  6.  three examples of such behavior are shown. 
The  simulation time is rescaled by multiplication by average particle flux, arriving at the step site, i.e. by $F d$. This is the mean distance $tFd$ that average step moves at unit time. All curves showing the system time evolution can be divided into three periods. In the first, very short phase of evolution  steps, initially ideally smooth, become rough. This phase is so short that it could be hardly detected in the plots shown in Figure 6. Microscopic step roughness depends on the temperature and on the particle flux $F$. After this stage, the step amplitude increases up to some limiting value at which the step pattern becomes stationary. The amplitude value at which meandering saturates depends mainly on the width of terraces $d$ being relatively weekly dependent on the other parameters. Saturation of step meandering is an entirely new phenomenon. It is observed only for regular meander patterns, having wavelengths longer than the terrace width $d$. Such behavior has not been predicted by the equations published in the literature, describing particle flux balance \cite{[23]}, nor seen  in previously  simulated systems \cite{P-L}. Evolution of systems simulated in Ref. \cite{P-L} is shown for much shorter times and higher fluxes than in our case, so they could rather classify as systems of shorter wavelengths. More important difference however is that the system there was simulated for infinite Schwoebel barrier, but overhangs and voids were excluded in the step configuration, what causes that the simulated systems behave in different way. 
 
Described here step pattern formation happens during crystal growth  and several different process play their role. A constant flux of particles $F$ arrives from above, subsequently distributed over the surface by the diffusion jumps. The particles attach to the steps, causing their movement forward  and some of them can detach from the steps, increasing particle density at the terrace. Very important element of the pattern formation is the existence of the diffusion along and across steps, usually faster and slower than the diffusion over the surface. The particle exchange between different terraces can be blocked by high Schwoebel barrier replacing regular meander structure by rough surface of irregular step forms (see Fig 2.). It is evident that the particle flow across steps is very important component  of the pattern formation. Comparing steps obtained for low and for high Schwoebel barrier, shown in Fig 2 we conclude that the meanders build up and order in regular structure all over whole system for communicating terraces, while the lack of inter--terrace communication destroys meanders  and the system becomes rough (chaotic). It is evident that the communication between steps is crucial factor for the analysis of the step dynamics.
 
Jumps of individual particles depend on the local surrounding at the surface, summing up and averaging into various fluxes, which are described and discussed on using different approaches \cite{[1],[2],[3],[4],[5],[22],[23],[27],[28],[29],[30]}.  The phenomena, which should be considered, depend on the analyzed  phase  of the growth.     
Below we describe the scenario, according to which the regular meanders are formed and they grow less or more quickly, depending on the difference between fluxes incoming towards step portion bulged towards upper or lower terrace. Because particles always attach step from one side, the step is asymmetric and particles coming to step from the terrace to the fragment exposed to the terrace have more space to attach than those, which arrive at the step bent unfavorably. Note, that we analyze situation of low or vanishing Schwoebel barrier, so the area from which particles arrive to the step is the same or similar for both step curvatures. However for the high step deformations, the difference of the number of adsorption sites density at the step for the positive curvature at meander top and the negative at the bottom, is significant and increases with the step bending. It is given by a product of  the particle flux $F d$, incoming to the step and  the difference of the number of sites at the concave and convex side of step divided by the step length. This last quantity can be expressed by the step curvature $\kappa=-d^2 z/dx^2=W/(2 \pi \lambda)^2$ and the lattice constant $a$, where $z(x)$ denotes the step shape. Finally, the flux difference at the both, concave and convex tips divided per unit  length  is 
\begin{equation}
j_{+} \sim    F d a \frac{d^2 z}{dx^2}
\end{equation}

Now, the flux coming out from the step is proportional to the number of particles, which are emitted, and to the current of diffusing particles. Step, which is concave emits more particles than the convex, and the difference is larger for larger curvature. The step stiffness is given by constant $\Gamma$. Particles desorb from the step and then they diffuse out. Stream of diffusing particles transports them to the another part of step, the process that is alimenting step dynamics. This flux goes mainly across steps, because this is the shortest path from one part of the  terrace  to another. Step communicate one with another by the exchange of particles. As we have discussed previously, their communication appears to be very important for regular meander pattern formation. It is evidently very important mechanism of the particle transport and it has to be proportional to the step length. Step length  grows with meander amplitude $W$ as $\sqrt{1+(2W/\lambda)^2}$ multiplied by a constant, dependent on the meander shape. Finally the following expression describes the outcoming  particle stream
\begin{equation}
\label{j-}
j_{-} \sim \Gamma \frac{d^2 z}{dx^2} \sqrt{1+(\frac{2W}{\lambda})^2}
\end{equation}
In addition to these two, specified above, main factors affecting the step dynamics, there are many others \cite{[23]},  which we did not  include here.  For example when one part of the step grows faster than the other, this causes that  particle density becomes locally lower, whereas in the surrounding regions, where step moves slowly, a net particle density is increased by all those, which are not built into the step. When particle density on top of one terrace surface region is higher than on the other, that difference in  particle density induces diffusion flux towards less dense regions which reduces  particle stream (\ref{j-}) a little. Such effect can be account for by correction of the constant in front this part of stream. 
In conclusion it seems  that in the case of regular and highly curved meander pattern these two expressions above, describe qualitatively all main aspects of step kinetics. 

When we add both terms, the  resulting equation is as follows
\begin{equation}
\label{flow}
\frac{d W}{dt}=\alpha F d \frac{W}{\lambda^2}-\kappa \Gamma \frac{D_0}{d} \frac{W}{\lambda^2} \sqrt{1+(\frac{2W}{\lambda})^2}
\end{equation}
 with the parameters $\alpha$ and $\kappa$ being generally constant, depending on several mechanisms of particle flow between  the step bent upward and step bent downward.
Equation (\ref{flow}) evidently sets the limit value of the meander amplitude $W_0$ , for which the step becomes stationary. It is given by
\begin{equation}
\label{W0}
W_0= \frac{\lambda}{2} \sqrt{\frac{F^2 d^4 \alpha ^2}{{D_0}^2 \kappa^2}-1}
\end{equation}
and is dependent on one constant $\alpha / \kappa$. Relation (\ref{W0}) fits saturation amplitudes of meanders in the simulated systems when constant  $\alpha / \kappa =3^. 10^4$.

Equation (\ref{flow}) can also be easily solved, describing the time evolution of the step meander amplitude, which has characteristic functional shape and which, for larger $W$ value is given by the hyperbolic tangent. In Figure 5  we have plotted time dependence of the meander amplitude $W$ for three different processes. We also fitted hyperbolic tangent shape to each evolution obtaining very good agreement, independently of the terrace width $d$ or the meander length $\lambda$. 

\section{Conclusions}
Kinetic MC simulations of GaN(0001) surface model were performed at different conditions. Meanders emerge out from initially straight steps and  for  wide range of parameters they grow up to given width and then  keep stable shape of well ordered waves during further crystal growing. Meander wavelength depends on the particle flux incoming to the step as $(Fd)^{-0.5}$. When wavelength is of the order of terrace width or lower, all meanders grow independently and eventually surface ends up as a rough structure. Larger meander wavelength create regular, ordered structure. The amplitude of meanders of this structure grows up to the limit value and then the step shape do not change during next stages of crystal growing. Such surface patterns are seen in the experiment as characteristic regular microscopic structures perpendicular to the initial step direction. We have shown that the crucial role in the explanation of described phenomena plays particle exchange between terraces via step. Such step to step communication is possible only when  the Schwoebel barrier is low enough. We described the mechanism of the step amplitude saturation  leading to stationary motion of the meandered step pattern.

\section{Acknowledgement}
 Research supported by the European Union within European Regional Development Fund, through grant Innovative Economy (POIG.01.01.02-00-008/08)


\bibliographystyle{elsarticle-num}

\end{document}